\documentclass[11pt]{elsart}
\usepackage{amssymb,amsmath}
\usepackage{graphicx}
\journal{Physica A}
\begin{document}
\begin{frontmatter}
\title{Long lasting instabilities in granular mixtures}
\author{H.Caps, R.Michel, N.Lecocq \and N.Vandewalle}           
\address{GRASP, Institut de Physique B5, University of Li\`ege, B-4000 Li\`ege, Belgium.}
\date{Received: 7 March 2003 / Revised version: date}
\begin{abstract}
We have performed experiments of axial segregation in the Oyama's drum. We have tested binary granular mixtures during very long times. The segregation patterns have been captured by a CCD camera and spatio-temporal graphs are created. We report the occurence of instabilities which can last several hours! We stress that those instabilities originate from the competition between axial and radial segregations. We put into evidence the occurence of giant fluctuations in the fraction of grain species along the surface during the unstable periods.
\end{abstract}

\end{frontmatter}
\section{Introduction}
\label{intro}

The most spectacular phenomenon associated with granular matter is certainly the phase segregation that one observes when a granular mixture is shaked, poured or rotated \cite{ristow}. Respectively, stripe \cite{demixing}, axial \cite{hill} and radial \cite{rajchenbach,bideau} segregation patterns have been reported. Recent experiments \cite{lecocq} and theoretical models \cite{boutreux} have been extensively performed in the particular case of stripe segregation. 

However, the least studied type of segregation is the case of the axial segregation in the so-called Oyama's drum \cite{hill,oyama}. The experiment consists in rotating a tube which is half filled with a sand mixture. The mixture is composed by 2 granular species differing in size and having different angles of repose. Due to the rotation, avalanches occur in the tube. Bands of the different granular species appear after a few minutes. A sketch and a picture of the axial segregation are illustrated in Figure 1. It has been reported \cite{hill} that the bands merge and only three stripes remain after very long times. One should also remark that Magnetic Resonance Imaging (MRI) experiments have shown that a radial segregation is also present in the center of the tube \cite{mri}. Small grains being located in the center of the tube.

\begin{figure}
\begin{center}
\resizebox{0.80\columnwidth}{!}{\includegraphics{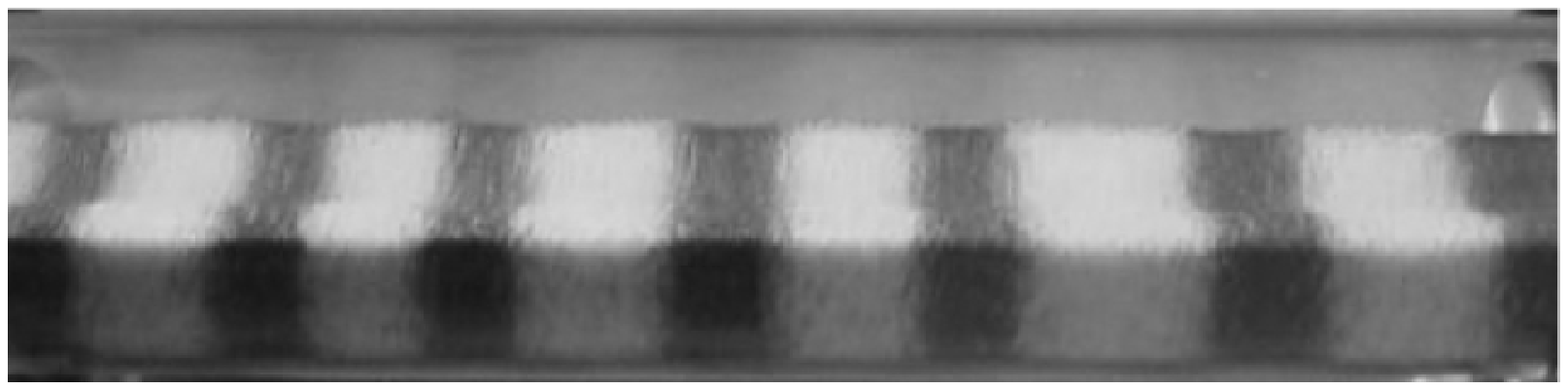}}
\vskip 0.3cm
\resizebox{0.80\columnwidth}{!}{\includegraphics{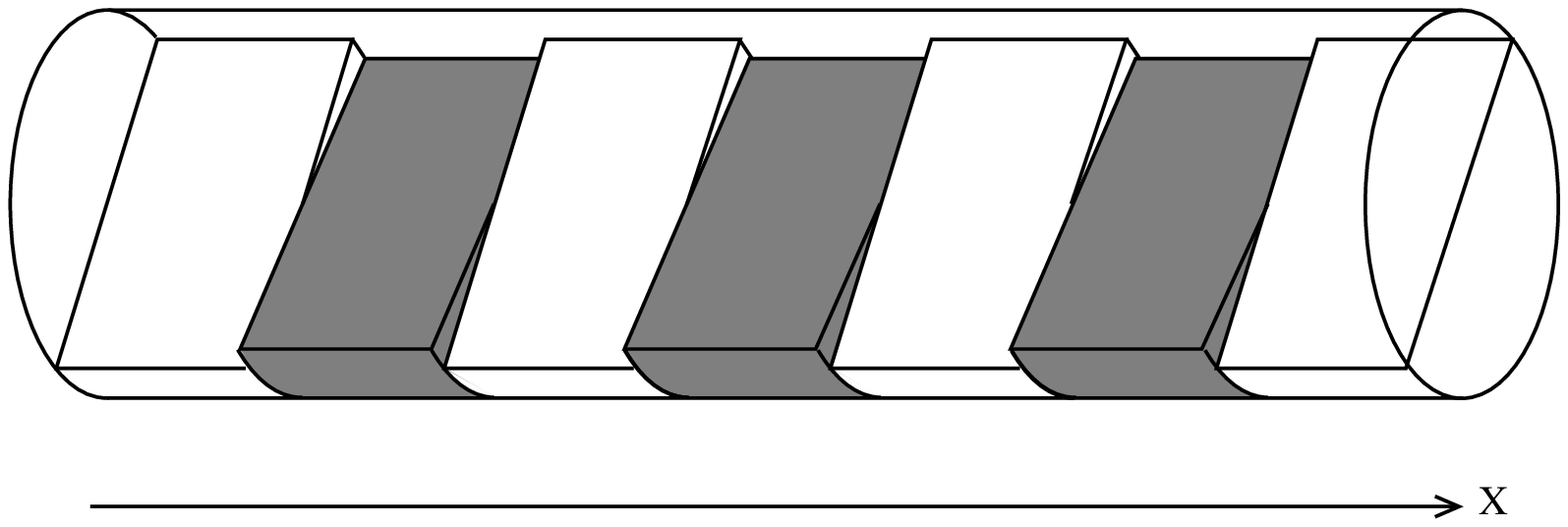}}
\end{center}
\caption{(top) Typical view of an axial segregation pattern. Dark grains are smaller than white grains. The picture emphasizes that bands are characterized by different slopes. (bottom) Sketch of the segregation pattern. The slope difference $\Delta \theta$ between adjacent bands leads to a segregation flux $\phi_s$ in the Savage's model.}
\label{fig:1}
\end{figure}

On the theoretical side, it has been proposed \cite{zik} that the axial segregation is similar to a spinodal decomposition. A simple and elegant model for diffusion of the different sand species along the tube axis has also been proposed by Savage \cite{savage}. This model considers the difference $\Delta \theta$ of dynamical angles for both species as a relevant parameter. If $\rho(x,t)$ is the density of one species along the tube axis $x$, the different slopes involve a segregation flux 
\begin{equation}
\phi_s = \beta {\partial \rho \over \partial x}
\end{equation} for that species. The positive coefficient $\beta$ is a function of the slope difference $\Delta \theta$ and the rotation speed $\omega$ of the tube. In addition, a Fickian diffusion 
\begin{equation}
\phi_d = - D {\partial \rho \over \partial x}
\end{equation} is taken into account. The diffusion coefficient $D$ is a function of the rotation speed $\omega$. Summing both fluxes and taking into account the continuity equation for $\phi_s+\phi_d$, one \cite{savage} obtains the classical diffusion equation
\begin{equation}
{\partial \rho \over \partial t} = D_e {\partial^2 \rho \over \partial x^2}
\end{equation} with an effective diffusion constant $D_e=D-\beta$ which could be negative for defined $\Delta \theta$ and $\omega$ values. A negative coefficient explains the formation of bands because any inhomogeneity in $\rho$ is then amplified towards full segregation. The Savage's model is a first step in the understanding of the phenomenon. The most elaborated model remains the one proposed by Zik and coworkers \cite{zik}. In the Zik's model, $\beta$ scales as $\beta \sim \rho (1-\rho)$ and the occurence of bands is attributed to a spinodal decomposition instead of a simple diffusion. More recently, molecular dynamics simulations have also been performed \cite{rapaport} in order to study the effect of friction. Axial segregation can be simulated with thousand of virtual grains.

We have performed new experiments in the Oyama's drum. We have produced axial segregation patterns and studied the phenomenon during several hours/days. In some experiments, we have obtained spectacular patterns that we call {\it instabilities}. The present paper deals with such unusual patterns. They are discussed in the light of previous theoretical arguments and earlier experiments. Mechanisms for instabilities appearance are proposed.

\section{Experiments}
\label{sec:1}

Our experimental setup is the following. A glass tube (diameter $40$ mm, length $300$ mm) is half filled with a binary sand mixture. The tube is rotated at a constant angular frequency $\omega$ which can be controlled. The rotation speed can range from 20 rotations per minute (rpm) to 40 rpm. The tube in rotation is perfectly horizontal. Two different sands have been selected for our study. For each type of sand, the granulometry has been selected by sifting. White grains have a mean diameter around $d=0.32$ mm. Black grains have a mean diameter around $d=0.22$ mm. The angle of repose for dark grains is $\theta=28^o\pm 1^o$. The angle of repose for the white grains is $\theta=33^o\pm 1^o$. Repose angles of different species have been measured by averaging the slope of a stable pile of the considered species along its surface. A CCD camera captures the pattern along the tube axis every minute. A computer collects the successive pictures. Each picture of the glass tube is rescaled into a line of pixels. This line represents the average density of the sand species along the $x$ axis. The successive lines of pixels are numerically glued together in order to form a space-time diagram of the evolving segregation pattern.

\begin{figure}
\begin{center}
\resizebox{4cm}{15cm}{\includegraphics{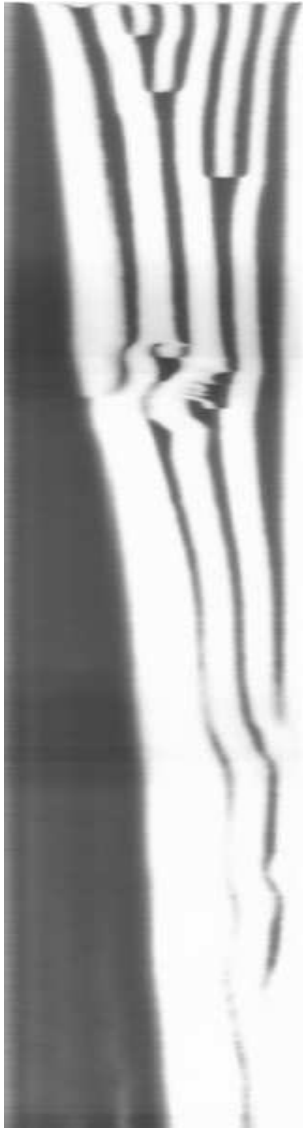}}
\end{center}
\caption{Spatio-temporal graph built with successive images of the Oyama's drum. Time goes down. The image width is 300 mm. The vertical axis corresponds to 64 hours of rotation at 40 rpm. White and dark bands are seen to merge towards full segregation of both granular species.}
\label{fig:2}
\end{figure}

Figure 2 presents a typical spatio-temporal graph built with 3840 images of the pattern, for a total period of 64 hours. In the picture, time goes down. The segregation pattern emerges within a few minutes. After a long time (several hours), the bands begin to merge. The merging of adjacent bands is rare. However, such an event is a rapid process (typically less than one minute) since the end part of a vanishing band is always abrupt. See three vanishing bands in the top of Figure 2. The number $N$ of bands seems to decrease exponentially with time. Figure 3 emphasizes such a decay for the particular data of Figure 2. We have fitted those data with the stretched exponential
\begin{equation}
N = 2 + (N_0-2) \exp \left( -\sqrt{t \over \tau} \right)
\end{equation} with $N_0$ being the initial number of bands and $\tau$ being a characteristic time for the decay. The data for others space-time graphs obtained with our setup are also well fitted with this empirical law. A systematic study of that particular dynamics is under way and will be published elsewhere \cite{renaud}.  Before looking at the details, we can make 3 general observations: 

(1) The segregation begins first at the extremities of the tube and then nucleates randomly in the tube. The spreading pattern formation is emphasized in Figure 4. The presence of tube ends (walls) affects locally the friction and induces the formation of bands at the extremities. We have noticed that at low rotation speeds ($\approx 10$ rpm), bands appear only at the extremities while the middle part of the tube is still mixed, even for very long times. 

(2) The topography of the granular surface is deeply affected by the segregation pattern. A close look at the surface reveals different slopes for the different sand species (see the top picture of Figure 1). Those slopes correspond to the dynamical angles of the species, which are $2^o$ larger than the repose angles. This is the main ingredient of both Savage \cite{savage} and Zik \cite{zik} models. 

(3) For very long times, the pattern tends to a complete segregation of both granular species. Only two bands are observed. This complete segregation is clearly observed at the end of Figure 2. One should remark that Frette and Stavans \cite{frette} did not always obtain a complete segregation pattern even after having performed very long experiences (but at a lower speed than in our cases).

\begin{figure}
\begin{center}
\resizebox{0.70\columnwidth}{!}{\includegraphics{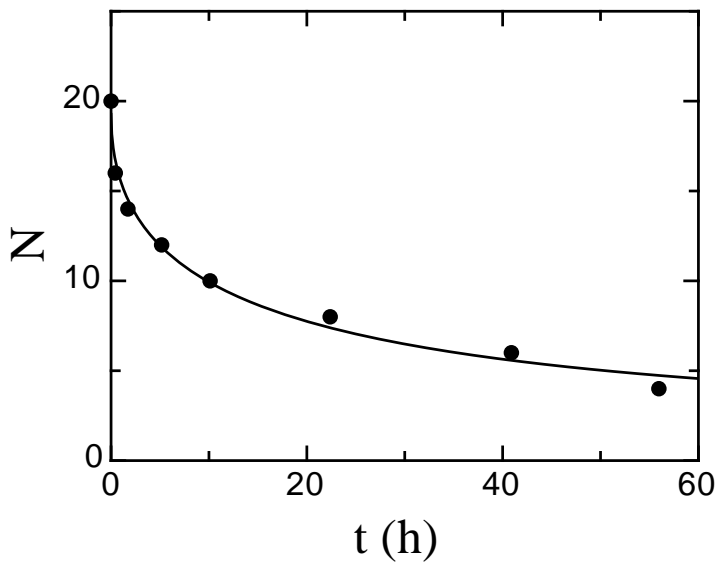}}
\end{center}
\caption{Decay of the number $N$ of bands as a function of time $t$ for the pattern of Figure 2. The curve is a fit using the empirical law of Eq.(4) with $N_0=20$ and $\tau=17 \,h$.}
\label{fig:3}
\end{figure}

\begin{figure}
\begin{center}
\resizebox{0.70\columnwidth}{!}{\includegraphics{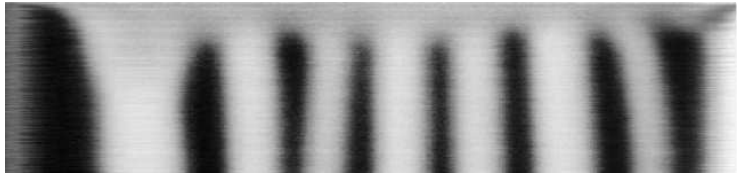}}
\end{center}
\caption{Pattern formation in the same conditions as Figure 2. Time goes down. The vertical axis corresponds to 30 minutes. For emphasizing the formation of bands, a picture has been taken every 10 seconds. The bands appear first at the extremities.}
\label{fig:4}
\end{figure}

In the great majority of our experiments, the successive merging of bands towards a full segregation of both granular species have been observed. However, looking at the details of Figure 2, one observes a turbulent-like pattern at the interface between white and dark bands. This particular event occured after 18 hours of rotation. The turbulent pattern  is emphasized in Figure 5. The duration of that event lasted about 5 hours. The instability of white/dark interfaces began when a new dark band nucleated in the middle part of a white band. This nucleation process involved the development of secondary bands in the neighborhood. Extremely slow undulations of the borders are also seen in the bottom of Figure 2, a long time after the `turbulent' event emphasized in Figure 5. 

The occurrence of such behavior is still mysterious. We did not discover the precise conditions on which they appear and disappear. However, it should be remarked that the occurence of such patterns is not an artefact. Indeed, they appeared in different experiments under various conditions, i.e. at various rotation speeds. Moreover, initial mixtures were premixed as homogeneously as possible.

\begin{figure}
\begin{center}
\resizebox{0.55\columnwidth}{!}{\includegraphics{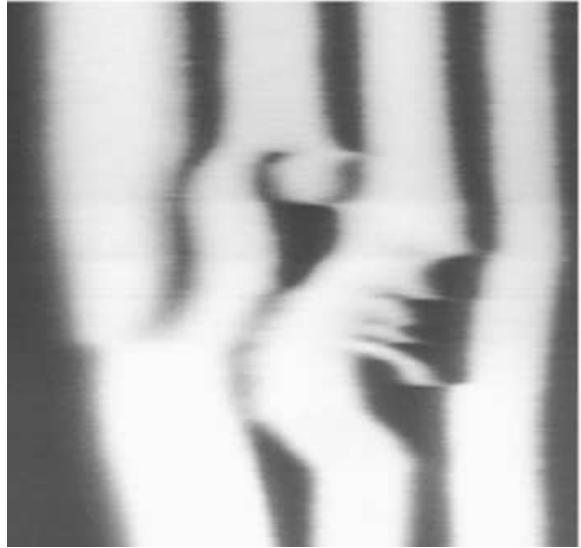}}
\end{center}
\caption{Blowup of the center part of Figure 2 emphasizing a `turbulent' event. Time goes down. The vertical axis correponds to a total of 10 hours. A thin dark band is seen to nucleate and to perturb neighboring bands.}
\label{fig:5}
\end{figure}

Figures 6 and 7 present additional patterns that we have also obtained. In those examples, the instability does not affect the total number of bands along the tube axis. In Figure 5, one observes undulations of the borders of white bands. In fact, thin dark bands nucleate periodically in the center of white bands before being expelled. In Figure 6, the situation is slightly different: two small bands are seen to detach from a white band and to travel from that band towards the next one.

\begin{figure}
\begin{center}
\resizebox{0.55\columnwidth}{!}{
\includegraphics{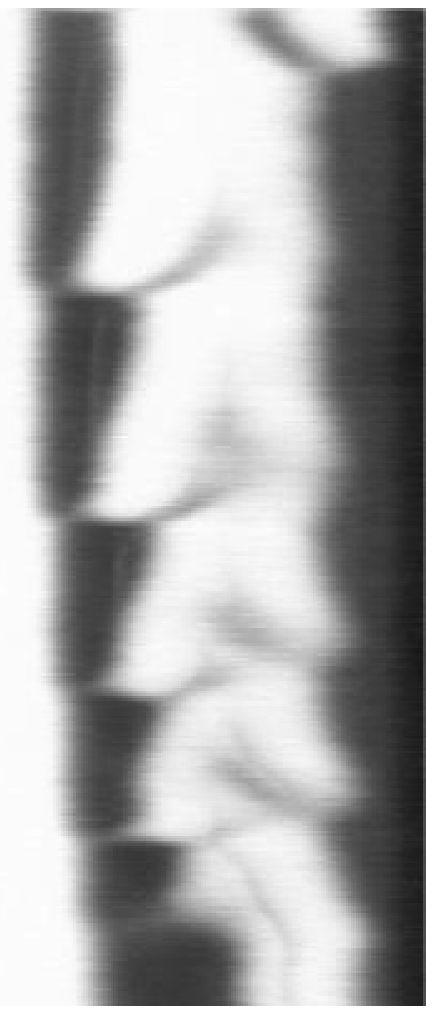}
\includegraphics{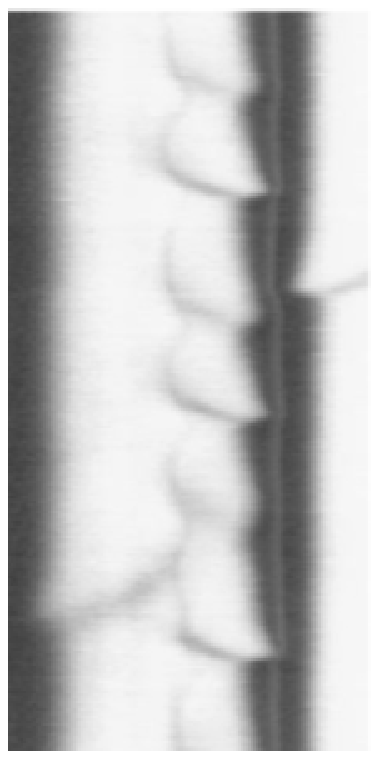}
}
\end{center}
\caption{Two typical spatio-temporal graphs of long lasting instabilities in a tube rotating at 26 rpm. Time goes down. The vertical axis corresponds to a total of 5h. Each image width is 50 mm.}
\label{fig:6}
\end{figure}

\begin{figure}
\begin{center}
\resizebox{0.5\columnwidth}{!}{
\includegraphics{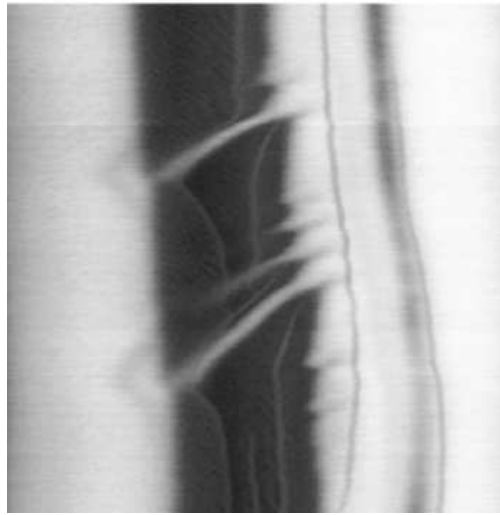}
}
\end{center}
\caption{Spatio-temporal graph of an other type of  instability. Time goes down. The vertical axis corresponds to a total of 3h, with a rotation speed of 38 rpm. The image width is 50 mm.}
\label{fig:7}
\end{figure}

Some instabilities are provoked by the nucleation of thin dark bands (see Figures 5 and 6). Others appear at the border of apparently stable white bands (see Figure 7). We will discuss the occurence of that phenomena in the next section.

\section{Discussion}

It should be noted that an earlier experimental work \cite{choo,choo2} put into evidence some travelling waves in the tube. This behavior was present for higher filling fraction of the tube ($> 55$\%) than in our experiments. Such travelling waves were associated with the early stages of the segregation (about 20 minutes), when inhomogeneities of the initial mixture provoke the nucleation and the motion of bands. Herein, the instabilities are seen to occur after very long times and last several hours. The nature of the instabilities could be quite different from those reported in \cite{choo,choo2}.
 
As mentionned at the beginning of this paper, MRI experiments \cite{mri} have put into evidence that radial segregation is always present. Small grains (the dark species herein) tend to segregate in the bulk. This implies that in our experiments, the white bands may have a dark core. We assist thus at the competition between axial and radial segregation in that regions. Any fluctuation of the radial pattern may provoke the nucleation of a thin dark band. Thereafter, this thin band travels slowly towards one border to be expelled. This coupling between surface and volume compositions could explain the instabilities. 

Surface and bulk processes are characterized by different time scales. Axial segregation has a larger characteristic time scale than radial segregation. This rate difference could also explain why our instabilities last several hours: the phenomenon comes from bulk fluctuations. MRI experiments could corroborate those mechanisms. 

Consider the fraction of dark species $x$ observed along the tube, measured by image analysis. Figure 8 presents the time evolution of the ratio
\begin{equation}
r = {x \over 1-x}
\end{equation} in the long experiment of Figure 5. The ratio $r$ emphasizes the surface `equilibrium' between both species along the tube. During the whole process, the ratio fluctuates around a linear trend close to $r=1$. However, during a period of instability, we observe giant fluctuations of $r$ up to $20\%$. This means that the compositions of both bulk and surface deviate strongly from equilibrium composition.

\begin{figure}
\begin{center}
\resizebox{0.90\columnwidth}{!}{\includegraphics{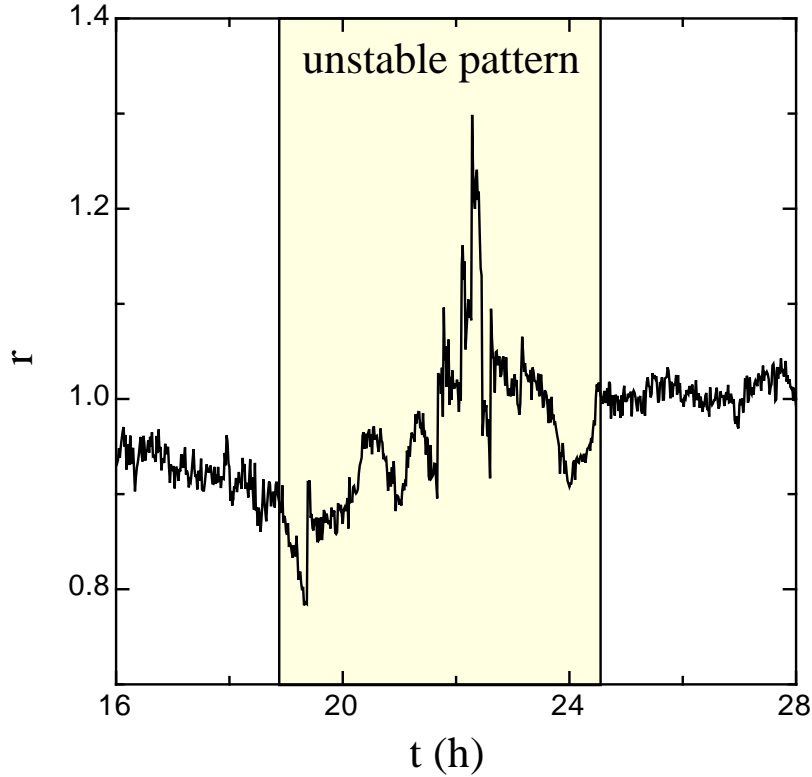}}
\end{center}
\caption{Evolution of the ratio of species fractions in the graph of Figure 5. The ratio $r$ fluctuate around a linear trend when bands are stable. Giant fluctuations are however seen when the pattern is unstable.}
\label{fig:8}
\end{figure}

\section{Summary}

Bulk segregation should be taken into account when axial patterns are discussed. Oyama's drum experiment is an example of 3-dimensional segregation. This has been already suggested by Hill et al \cite{mri}. Moreover, we have reported hereabove additional effects such as long lasting instabilities which arise from bulk processes. The dynamics of those patterns are extremely slow with respect to the formation/vanishing of the bands. We stress that those instabilities originate from the competition between {\it axial} and {\it radial} segregations. As a consequence, the Savage model, which considers only surface effects should be extended to take into account the bulk distribution of granular species.

The physical conditions, such as rotation speed and granulometry, which imply the formation of instabilities, are still unclear. This point needs a systematic experimental study. This represents a huge amount of work since each experiment lasts several hours.

\section*{Acknowledgements}
HC is financially supported by the FRIA (Brussels, Belgium). This work is also suppported through the ARC contract n$^\circ$02/07-293. The authors would also like to thank M. Ausloos for fruitful discussions.
%%%%%%%%%%%%%%%%%%%%%%%%

\end{document}